\documentclass[natbib209]{emulateapj}
\usepackage{graphicx}
\usepackage{rotating}
\usepackage[usenames]{color}
\usepackage[colorlinks=true, pdfstartview=FitV, linkcolor=OrangeRed, citecolor=blue, urlcolor= red]{hyperref}

\shortauthors{Oh et al.}
\shorttitle{The central slope of dark matter cores in dwarf galaxies: Simulations vs. THINGS}

\begin{document}
\bibliographystyle{astroads}

\renewcommand{\thefootnote}{\fnsymbol{footnote}}
\newcommand{\C}{\ensuremath{\mathfrak{C}}}
\newcommand{\VNFW}{\ensuremath{V_{200}}}
\newcommand{\Vh}{\ensuremath{V_h}}
\newcommand{\kms}{\ensuremath{\mathrm{km}\,\mathrm{s}^{-1}}}
\newcommand{\cm}{\ensuremath{\mathrm{cm}^{-2}}}
\newcommand{\kmsnospace}{\ensuremath{\mathrm{km}\,\mathrm{s}^{-1}}}
\newcommand{\barkms}{\ensuremath{\mathrm{km}\,\mathrm{s}^{-1}\,\mathrm{kpc}^{-1}}}
\newcommand{\kmsMpc}{\ensuremath{\mathrm{km}\,\mathrm{s}^{-1}\,\mathrm{Mpc}^{-1}}}
\newcommand{\etal}{et al.}
\newcommand{\LCDM}{$\Lambda$CDM}
\newcommand{\ML}{\ensuremath{\Upsilon_{\star}}}
\newcommand{\MLfix}{\ensuremath{\Upsilon_{\star}^{fix}}}
\newcommand{\MLfree}{\ensuremath{\Upsilon_{\star}^{\rm free}}}
\newcommand{\MLsps}{\ensuremath{\Upsilon_{\star}^{3.6}}}
\newcommand{\MLspl}{\ensuremath{\Upsilon_{\star}^{4.5}}}
\newcommand{\MLk}{\ensuremath{\Upsilon_{\star}^{K}}}
\newcommand{\MLb}{\ensuremath{\Upsilon_{\star}^{B}}}
\newcommand{\MLmax}{\ensuremath{\Upsilon_{max}}}
\newcommand{\Lsun}{\ensuremath{\rm{{L}_{\odot}}}}
\newcommand{\Msun}{\ensuremath{\rm{{M}_{\odot}}}}
\newcommand{\mass}{\ensuremath{\rm{{\cal M}}}}
\newcommand{\magsq}{\ensuremath{\mathrm{mag}\,\mathrm{arcsec}^{-2}}}
\newcommand{\Lsundens}{\ensuremath{\rm L_{\odot}\,\mathrm{pc}^{-2}}}
\newcommand{\surfdens}{\ensuremath{\rm M_{\odot}\,\mathrm{pc}^{-2}}}
\newcommand{\cubedens}{\ensuremath{M_{\odot}\,\mathrm{pc}^{-3}}}

\long\def\Ignore#1{\relax}
\long\def\Comment#1{{\footnotesize #1}}


\title{The central slope of dark matter cores in dwarf galaxies: \\ Simulations vs. THINGS}

\author{Se-Heon Oh\altaffilmark{1,8},
Chris Brook\altaffilmark{2},
Fabio Governato\altaffilmark{3},
Elias Brinks\altaffilmark{4},
Lucio Mayer\altaffilmark{5},
W.J.G. de Blok\altaffilmark{1}, \\
Alyson Brooks\altaffilmark{6} and
Fabian Walter\altaffilmark{7}}

\email{seheon\_oh@ast.uct.ac.za}
\email{cbbrook@uclan.ac.uk}
\email{fabio@astro.washington.edu}
\email{E.Brinks@herts.ac.uk}
\email{lmayer@physik.unizh.ch}
\email{edeblok@ast.uct.ac.za}
\email{abrooks@tapir.caltech.edu}
\email{walter@mpia.de}

\altaffiltext{1}{Astronomy Department, Astrophysics, Cosmology and Gravity Centre (ACGC), University of Cape Town, Private Bag X3, Rondebosch 7701, South Africa}
\altaffiltext{2}{Jeremiah Horrocks Institute, University of Central Lancashire, Preston, Lancashire, PR1 2HE, United Kingdom}
\altaffiltext{3}{Astronomy Department, University of Washington, Seattle, Washington 98195, USA}
\altaffiltext{4}{Centre for Astrophysics Research, University of Hertfordshire, College Lane, Hatfield, AL10 9AB, United Kingdom}
\altaffiltext{5}{Institute for Theoretical Physics, University of Zurich, Winterthurestrasse 190, 8057 Z\"urich}
\altaffiltext{6}{Theoretical Astrophysics, California Institute of Technology, MC 350-17, Pasadena, California 91125, USA}
\altaffiltext{7}{Max-Planck-Institut f\"ur Astronomie, K\"onigstuhl 17, 69117 Heidelberg, Germany}
\altaffiltext{8}{Square Kilometre Array South African Fellow}


\begin{abstract}
We make a direct comparison of the derived dark matter (DM)
distributions between hydrodynamical simulations of dwarf
galaxies assuming a \LCDM\ cosmology and the observed dwarf
galaxies sample from the THINGS survey in terms of (1) the
rotation curve shape and (2) the logarithmic inner density slope
$\alpha$ of mass density profiles.
The simulations, which include the effect of baryonic feedback processes,
such as gas cooling, star formation, cosmic UV background heating and most
importantly physically motivated gas outflows driven by supernovae (SNe),
form bulgeless galaxies with DM cores. We show that the stellar and baryonic
mass is similar to that inferred from photometric and kinematic
methods for galaxies of similar circular velocity. Analyzing the
simulations in exactly the same way as the observational sample
allows us to address directly the so-called ``cusp/core'' problem
in the \LCDM\ model. We show that the rotation curves of the simulated dwarf galaxies
rise less steeply than CDM rotation curves and are consistent with those of the THINGS dwarf galaxies.
The mean value of the logarithmic inner density slopes
$\alpha$ of the simulated galaxies' dark matter density profiles is $\sim$$-0.4\pm0.1$, which
shows good agreement with $\alpha$$=$$-0.29\pm0.07$ of the THINGS
dwarf galaxies. The effect of non-circular motions is not significant
enough to affect the results. This confirms that the baryonic feedback
processes included in the simulations are efficiently able to
make the initial cusps with $\alpha\sim$$-1.0$ to $-1.5$ predicted by
dark-matter-only simulations shallower, and induce DM halos
with a central mass distribution similar to that observed in nearby
dwarf galaxies.
\end{abstract}

\section{Introduction} \label{Intro}
The dark matter (DM) distributions at the centers of galaxies have been intensively discussed
from both observational and theoretical sides for almost two decades ever since
high-resolution N-body dark matter simulations assuming a universe dominated by cold dark matter (CDM) and a
cosmological constant $\Lambda$ were performed.
The \LCDM\ simulations have invariably predicted a dark matter density distribution which diverges
toward the centers of galaxies
(\citeauthor{Moore_1994} \citeyear{Moore_1994};
Navarro, Frenk \& White \citeyear{NFW_1996}, \citeyear{NFW_1997};
\citeauthor{Moore_1999a} \citeyear{Moore_1999a};
\citeauthor{Ghigna_2000} \citeyear{Ghigna_2000};
\citeauthor{Klypin_2001} \citeyear{Klypin_2001};
\citeauthor{Power_2002} \citeyear{Power_2002};
\citeauthor{Stoehr_2003} \citeyear{Stoehr_2003};
\citeauthor{Navarro_2004} \citeyear{Navarro_2004};
\citeauthor{Reed_2005} \citeyear{Reed_2005};
\citeauthor{Diemand_2008} \citeyear{Diemand_2008}).
In order to describe such cusp--like dark matter distributions, Navarro,
Frenk and White (\citeyear{NFW_1995}; \citeyear{NFW_1996}) proposed a profile (hereafter NFW profile)
which can be approximated by two power laws, $\rho\sim$$r^{-1.0}$ and $\rho\sim$$r^{-3.0}$ to describe the inner and outer regions
of a dark matter halo, respectively.
In particular, the central cusp feature with $\rho$$\sim$$r^{-1.0}$ has provided a useful test for \LCDM\ cosmology and sparked interest
in seeking constraints by observing mass distributions at the center of galaxies
(e.g., \citeauthor{McGaugh_2003} \citeyear{McGaugh_2003}).
It is worth mentioning that recent simulations show shallower slopes with
$\alpha\sim$$-0.8$ at radii of 120 pc (\citeauthor{Stadel_2009} \citeyear{Stadel_2009})
although DM slopes at $r < 1$ kpc are generally steeper (see also \citeauthor{Navarro_2010} \citeyear{Navarro_2010}).

The prediction of a central cusp from the \LCDM\ model has
been seriously challenged by observations of dwarf and Low Surface
Brightness (LSB) disk galaxies.  Observations of dwarf and LSB galaxies generally
indicate a constant matter distribution toward their centers, with mass density profiles with
a kpc sized core radius. This discrepancy of central dark matter distributions in dwarf
galaxies in \LCDM\ simulations and observations is referred to as the
``cusp/core'' problem. This is a fundamental problem for \LCDM\, together with the likely connected
substructure and angular momentum problems
(\citeauthor{Moore_1999} \citeyear{Moore_1999};
\citeauthor{Klypin_1999} \citeyear{Klypin_1999};
\citeauthor{Simon_2007} \citeyear{Simon_2007};
\citeauthor{Navarro_White_1994} \citeyear{Navarro_White_1994};
\citeauthor{Dutton_2009} \citeyear{Dutton_2009}).

Compared to other types of galaxies, dwarf galaxies provide us with a good opportunity for measuring the
dark matter distribution near the centers of galaxies due to the fact that they have a simple dynamical
structure (disk galaxies without bulges) but also have low baryon fractions and hence less of a
dynamical contribution by baryons (\citeauthor{deBlok_1997} \citeyear{deBlok_1997}).
Recently, \cite{Oh_2011} presented high-resolution dark matter
density profiles of 7 dwarf galaxies taken from ``The H{\sc i} Nearby Galaxy Survey''
(THINGS; \citeauthor{Walter_2008} \citeyear{Walter_2008}). The high quality data from THINGS
significantly minimizes observational uncertainties and thus allows us to investigate
the central dark matter distribution of the dwarf galaxies in detail.
Mass models of stars and gas are constructed
using the {\it Spitzer} IRAC 3.6$\mu$m data from the ``{\it Spitzer} Infrared Nearby Galaxies
Survey'' (SINGS; \citeauthor{Kennicutt_2003} \citeyear{Kennicutt_2003}) and the total integrated
THINGS H{\sc i} map, respectively. The kinematics of baryons are then subtracted
from the total kinematics of the galaxies in order to derive the dark matter distributions.

One of the main results of \cite{Oh_2011} was to robustly
confirm that the rotation curves of the 7 THINGS dwarf galaxies rise
too slowly to be consistent with a cusp feature at their centers. Moreover, the
mean value of the logarithmic inner slopes of the dark matter
density profiles is $\alpha=-0.29\pm0.07$, which significantly deviates from the $\alpha\sim$$-1.0$
predicted from dark-matter-only simulations. The exquisite data used in this study allowed
unprecedented treatment of the effects of observational uncertainties, such as beam smearing,
center offset and non-circular motions, which may play a role in hiding central cusps
(\citeauthor{Blais-Ouellette_1999} \citeyear{Blais-Ouellette_1999};
\citeauthor{van_den_Bosch_2000} \citeyear{van_den_Bosch_2000};
\citeauthor{Bolatto_2002} \citeyear{Bolatto_2002};
\citeauthor{Swaters_2003} \citeyear{Swaters_2003};
\citeauthor{Simon_2003} \citeyear{Simon_2003};
\citeauthor{Rhee_2004} \citeyear{Rhee_2004};
\citeauthor{Gentile_2005} \citeyear{Gentile_2005};
\citeauthor{Spekkens_2005} \citeyear{Spekkens_2005};
\citeauthor{Oh_2008} \citeyear{Oh_2008}).
The results have thus significantly strengthened the observational evidence that
the dark matter distribution near the centers of dwarf galaxies follows a near--constant density core.

Before using these results as a repudiation of cold dark matter, however, one must also
examine carefully the modeling on which the central cusp predictions are based,
which are usually done using N-body simulations that  include only the effects of gravity on structure
formation. Although baryons make up only $\sim$$14 \%$ of the matter of the Universe, this dissipative constituent
of the Universe cools with cosmological time and accumulates in the central regions of DM halos, making up a 
dynamically important fraction. Several mechanisms have been proposed whereby these central baryons can affect the
central cusp--like dark matter distribution which is found in pure dark matter simulations. A rapid
change in potential (faster than the dynamical time) due to star burst triggered outflows, is
one mechanism which has been shown to be capable of transforming cusp--like profiles into (flatter) cores
(\citeauthor{NFW_1996} \citeyear{NFW_1996};
\citeauthor{Read_2005} \citeyear{Read_2005}
although see \citeauthor{Gnedin_2002} \citeyear{Gnedin_2002}).
Supernova driven random bulk motions of gas in proto-galaxies
(\citeauthor{Mashchenko_2006} \citeyear{Mashchenko_2006})
has also been shown in models to flatten cusps, as have the effects of dynamical friction
acting on gas clumps (\citeauthor{ElZant_2001} \citeyear{ElZant_2001}), and the transfer of
angular momentum from baryons to the dark matter (\citeauthor{Tonini_2006} \citeyear{Tonini_2006}).
Modeling an inhomogeneous multi--phase, interstellar medium is critical for simulating the
baryonic feedback processes in galaxies (\citeauthor{Robertson_2008} \citeyear{Robertson_2008};
\citeauthor{Ceverino_2009} \citeyear{Ceverino_2009}). Yet cosmological simulations have not,
until recently, been able to achieve enough resolution to model even such an inhomogeneous
ISM, but have been forced to treat the important  processes of star formation and feedback
as ``sub-grid'' physics, averaging star formation and supernova feedback over large volumes,
compared to the typical structural scales ($<$1 kpc) of small galaxy disks.

Most recently, \cite{Governato_2010} have performed high-resolution
cosmological N--body+Smoothed Particle Hydrodynamic (SPH)
simulations of dwarf galaxies under the \LCDM\ paradigm, that include
the effect of baryonic feedback processes, such
as gas cooling, star formation, cosmic $\it UV$ background heating and
most importantly physically motivated gas outflows driven by SNe.
The major finding of \cite{Governato_2010} was that once star formation is
associated with high density gas regions, a significant amount of
baryons with low angular momentum is efficiently removed by strong
SNe--driven injection of thermal energy and the following gas outflows that carry an amount of gas at a rate of 2--6 times the
local star formation rate. The large scale outflows in turn induce two effects:
the loss of low angular momentum gas from the central regions prevents the formation
of bulges in low mass systems. 
Secondly, the clumpy nature of the gas and rapid ejection on short timescales has dynamical effects
on the dark matter potential, creating a shallower density profile
(see \citeauthor{Mashchenko_2008} \citeyear{Mashchenko_2008};
\citeauthor{Ceverino_2009} \citeyear{Ceverino_2009};
\citeauthor{Mo_2004} \citeyear{Mo_2004};
\citeauthor{Mashchenko_2006} \citeyear{Mashchenko_2006}).

Moreover, the simulated galaxies have a z=0 baryonic budget
consistent with photometric and kinematic estimates
(\citeauthor{van_den_Bosch_2001b} \citeyear{van_den_Bosch_2001b};
\citeauthor{McGaugh_2010} \citeyear{McGaugh_2010}).
The kinematic properties of the simulated dwarf galaxies are very similar to those
of the THINGS dwarf galaxies in terms of their maximum rotation velocity
($\sim$$60$ \kms) and dynamical mass ($\sim$$10^{9}$
$\ensuremath{M_{\odot}}$), allowing a direct comparison between the simulations and 
observations to be made.

\begin{figure*}
\epsscale{1.0}
\includegraphics[angle=0,width=1.0\textwidth,bb=40 195 570 555,clip=]{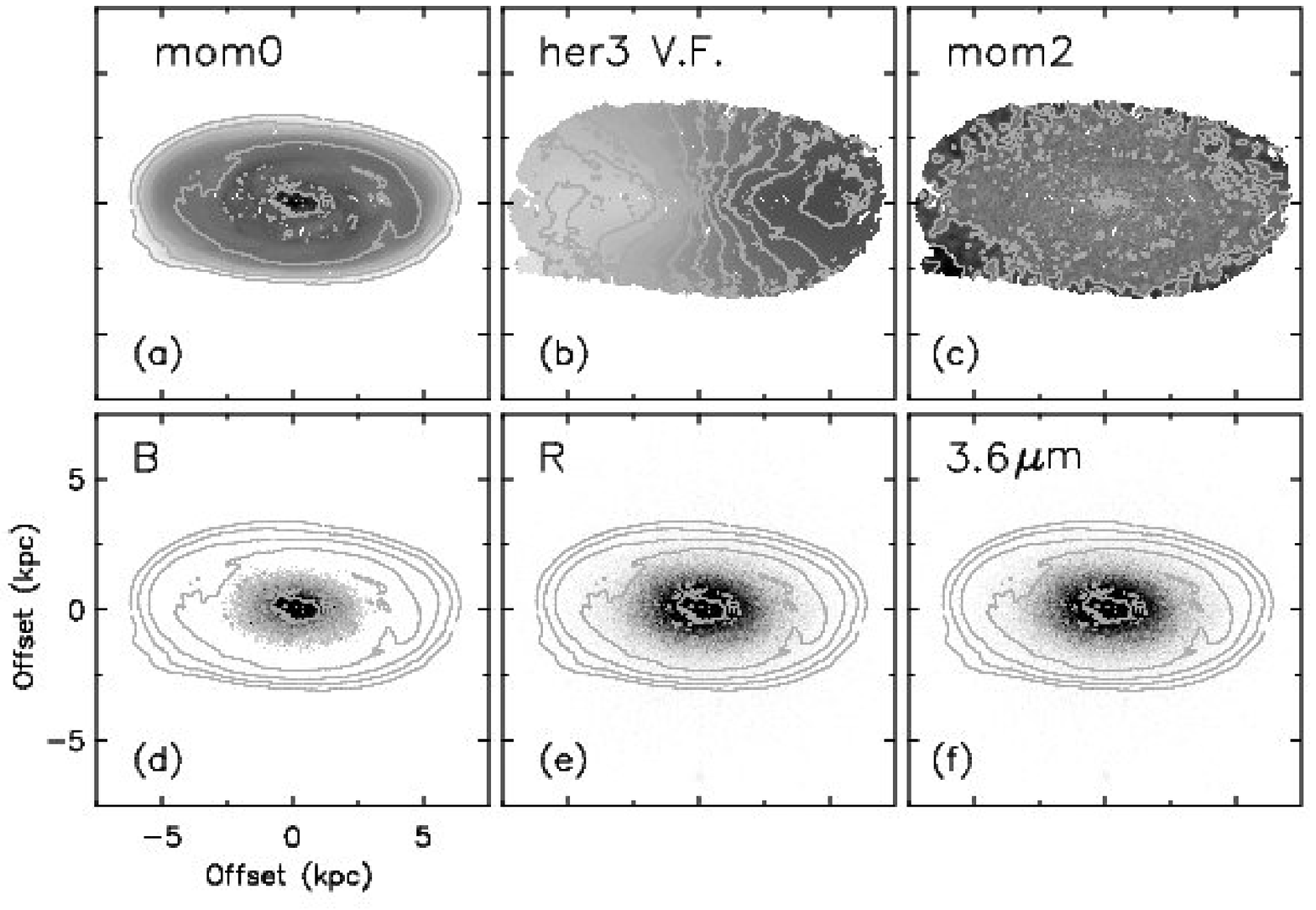}
\caption{\scriptsize Gas and stellar components of DG1.
{\bf (a):} Simulated integrated gas map (moment 0). Contours start from a column density of $\rm 10^{20}$\,\cm\ to $\rm 10^{22}$\,\cm\ in steps of $10^{0.4}$\,\cm.
{\bf (b):} Hermite $h_{3}$ velocity field. Velocity contours run from $-50$\,\kms\,to $50$\,\kms\,with a
spacing of $10$\,\kms.
{\bf (c):} Velocity dispersion map (moment 2). Velocity contours run from $5$\,\kms\,(gray) to $15$\,\kms\,(black) with a spacing of $4$\,\kms.
{\bf (d)(e)(f):}: Total intensity maps in simulated optical {\it B}, {\it R} and Spitzer IRAC 3.6 $\mu$m bands with superimposed
contours of the gas map in the panel (a).
\label{DG1_images}}
\end{figure*}

A first estimate of the bulge-to-disk ratio and DM properties of these models was presented in \cite{Governato_2010}
by fitting a Sersic profile to artificial {\it i--}band images and fitting a rotation curve to the rotational
motions of cold gas using a tilted-ring analysis. 
Here, we make more direct comparisons with observations, {\em applying the same
analysis techniques and tools} to the simulations as done in the most recent observational
sample of analoguous galaxies. The determination of the DM slopes for the simulations
has to be done in the same way as in observations since there are various aspects
that can affect the answer, such as beam smearing, center offset and non-circular motions,
which were not taken into account when the slope was derived from the raw simulation data in
\cite{Governato_2010}.
This approach will have two crucial benefits: (1) provide a strong test of theoretical
predictions and (2) test the extent of observational biases possibly
associated to current models of baryon/DM mass decompositions in
real galaxies, in particular the recovery of non-circular motions
and pressure support induced by SN feedback (\citeauthor{Valenzuela_2007} \citeyear{Valenzuela_2007};
\citeauthor{Dalcanton_2010} \citeyear{Dalcanton_2010}).

The structure of this paper is as follows. The simulations are described in Section~\ref{SPH_description}.
In Section~\ref{Cosmic_HDS}, we present the dark matter mass modeling of the simulated dwarf galaxies.
Section~\ref{Sim_THINGS} compares the derived dark matter distributions from the simulations
with those of the THINGS dwarf galaxies. Lastly, we summarize the main results of this paper and
conclusions in Section~\ref{conclusion}.

\section{The SPH Treecode Gasoline}\label{SPH_description}

The halos for these simulations  were selected from a set of large--scale, 
low--resolution, dark matter only simulation run in a concordance, flat,
$\Lambda$-dominated cosmology: $\Omega_0=0.24$, $\Lambda$=0.76,
$h=0.73$, $\sigma_8=0.77$, and $\Omega_{b}=0.042$ (\citeauthor{Verde_2003} \citeyear{Verde_2003}).
The size of the box, 25~Mpc, is large enough to provide
realistic torques for the small galaxies used in this work. The power
spectra to model the initial linear density field were calculated
using the CMBFAST code to generate transfer functions.
To include the effects of cosmic torques from the large scale
structure we used the volume renormalization (or ``zoom in'')
technique (\citeauthor{Katz_1993} \citeyear{Katz_1993}). Dark matter particle masses in the high
resolution regions are 1.6$\times$10$^4$ M$_{\odot}$, while the mass of star particles
is only 1000 M$_{\odot}$ and the force resolution, i.e., the gravitational
softening, is 86 pc. In total, at z$=$0 there are $3.3 \times 10^6$ particles within the virial radius
of the simulated dwarf galaxy, hereafter referred to as DG1. For all particle species, the gravitational spline
softening, $\epsilon(z)$, was evolved in a comoving manner from the
starting redshift (z $\sim$ 100) until z=8, and then remained fixed at
its final value from z=8 to the present.
At z $=$ 0 the virial masses of the halos that we studied in this paper
are 3.5 (DG1) and 2.0 (DG2) $\times$ 10$^{10}$ M$_{\odot}$ (the virial
mass is measured within the virial radius R$_{\rm vir}$, the radius
enclosing an overdensity of 100 times the cosmological critical
density). 

To evolve the simulations described here we have used the fully
parallel, N-body+SPH code GASOLINE
to compute the evolution of both the collisionless and dissipative
component in the simulations. A detailed description of the code is
available in the literature (\citeauthor{Wadsley_2004} \citeyear{Wadsley_2004}).
The version of the code used in this paper includes radiative cooling and accounts for the effect of
a uniform background radiation field on the ionization and excitation
state of the gas. The cosmic ultraviolet background is implemented
using the Haardt-Madau model (\citeauthor{Haardt_1996} \citeyear{Haardt_1996}), including photoionizing and
photoheating rates produced by Pop\,III stars, QSOs and galaxies
starting at $z=9$. We use a standard cooling function for a primordial
mixture of atomic hydrogen and helium at high gas temperatures and we
include low temperature cooling (\citeauthor{Mashchenko_2006} \citeyear{Mashchenko_2006}).

In the simulations described in this paper star formation occurs when
cold gas reaches a given threshold density (e.g., \citeauthor{Stinson_2006} \citeyear{Stinson_2006})
typical of actual star forming regions (we used 100 atomic--mass--unit (amu)\,cm$^{-3}$). Star formation (SF) then proceeds at a
rate proportional to $\rho_{gas}^{1.5}$, i.e. locally enforcing a
Schmidt law. The adopted feedback scheme is implemented by releasing
thermal energy from SNe into the gas surrounding each star
particle (\citeauthor{Stinson_2006} \citeyear{Stinson_2006}). The energy release rate is tied to the time of
formation of each particle (which effectively ages as a single stellar
population with a Kroupa IMF). To model the effect of feedback at
unresolved scales, the affected gas has its cooling shut off for a
time scale proportional to the Sedov solution of the blast wave
equation, which is set by the local density and temperature of the gas
and the amount of energy involved. The effect of feedback is to
regulate star formation in the discs of massive galaxies and to
greatly lower the star formation efficiency in galaxies with peak
circular velocity in the 50 $<$ V$_c$ $<$150 \kms\ range (\citeauthor{Brooks_2007} \citeyear{Brooks_2007}).
At even smaller halo masses (V$_c$ $<$ 20-40 \kms) the collapse of
baryons is largely suppressed by the cosmic UV field (\citeauthor{Quinn_1996} \citeyear{Quinn_1996};
\citeauthor{Okamoto_2008} \citeyear{Okamoto_2008}; \citeauthor{Gnedin_2010} \citeyear{Gnedin_2010}).

Other than the density threshold only two other parameters are needed, the
star formation efficiency ($\epsilon$SF = 0.1) and the fraction of SN
energy coupled to the ISM ($\epsilon$SN=0.4). 
The model galaxies studied in this paper are those published in \cite{Governato_2010}. We verified that
the addition of full metal cooling and the increase of $\epsilon$SN to 1 
does not substantially change the structural properties of the galaxy. However
the amount of stars formed decreases by 40\%.

As a benchmark against which the effects of baryonic feedback processes can be gauged, \cite{Governato_2010}
performed additional runs for analoguous model galaxies of DG1, called DG1DM and DG1LT. DG1DM uses the same initial
conditions as DG1 but includes only the dark matter component. It has a DM slope of $\alpha\sim$$-1.3$
similar to those found from similar simulations (\citeauthor{Springel_2008} \citeyear{Springel_2008}).
DG1LT is a version of DG1 using a lower density threshold (0.1 amu\,cm$^{-3}$) at a lower resolution
(the force resolution is $\sim$116 pc), and outflows in this model are negligible. DG1LT has a cusp and its
DM slope is similar to that of DG1DM ($\alpha\sim$$-1.3$).

To properly compare the outputs from the simulation to real galaxies
and make accurate estimates of the {\it observable} properties of
galaxies (e.g., \citeauthor{Sanchez-Janssen_2010} \citeyear{Sanchez-Janssen_2010}),
we used the Monte Carlo radiation transfer code
{\it{SUNRISE}} (\citeauthor{Jonsson_2010} \citeyear{Jonsson_2010}) to generate artificial optical images and
spectral energy distributions (SEDs) of the outputs of our run.
{\it{SUNRISE}} allows us to measure the dust reprocessed SED of every
resolution element of the simulated galaxies, from the far UV to the
far IR, with a full 3--D treatment of radiative transfer.
We place the simulated galaxies at an inclination of 60 degrees.
However, in applying our analysis tools, the inclination is considered
a free parameter, in keeping with the techniques applied to observed galaxies. 
Filters mimicking those of the SDSS survey are used to create mock
observations.

We note that in the runs adopting the ``high threshold'' SF, feedback
produces winds that are comparable in strength to those found in
real galaxies of similar mass. However, in our simulations the cold ISM is
still only moderately turbulent ($\sim$ 5--10 \kms\ at $\rm z=0$), consistent
with observations, and the galaxies match the observed stellar
and baryonic Tully--Fisher relation (\citeauthor{Governato_2009} \citeyear{Governato_2009}),
as the SF efficiency is regulated to form an amount of stars similar to that of real dwarf
galaxies of similar rotation velocity.

In the following sections, we perform dark matter mass modeling of both DG1 and DG2
in exactly the same way as the THINGS dwarf galaxies sample described in
\citeauthor{Oh_2011} (\citeyear{Oh_2008}, \citeyear{Oh_2011}).

\begin{figure*}
\epsscale{1.0}
\includegraphics[angle=0,width=1.0\textwidth,bb=30 175 550 685,clip=]{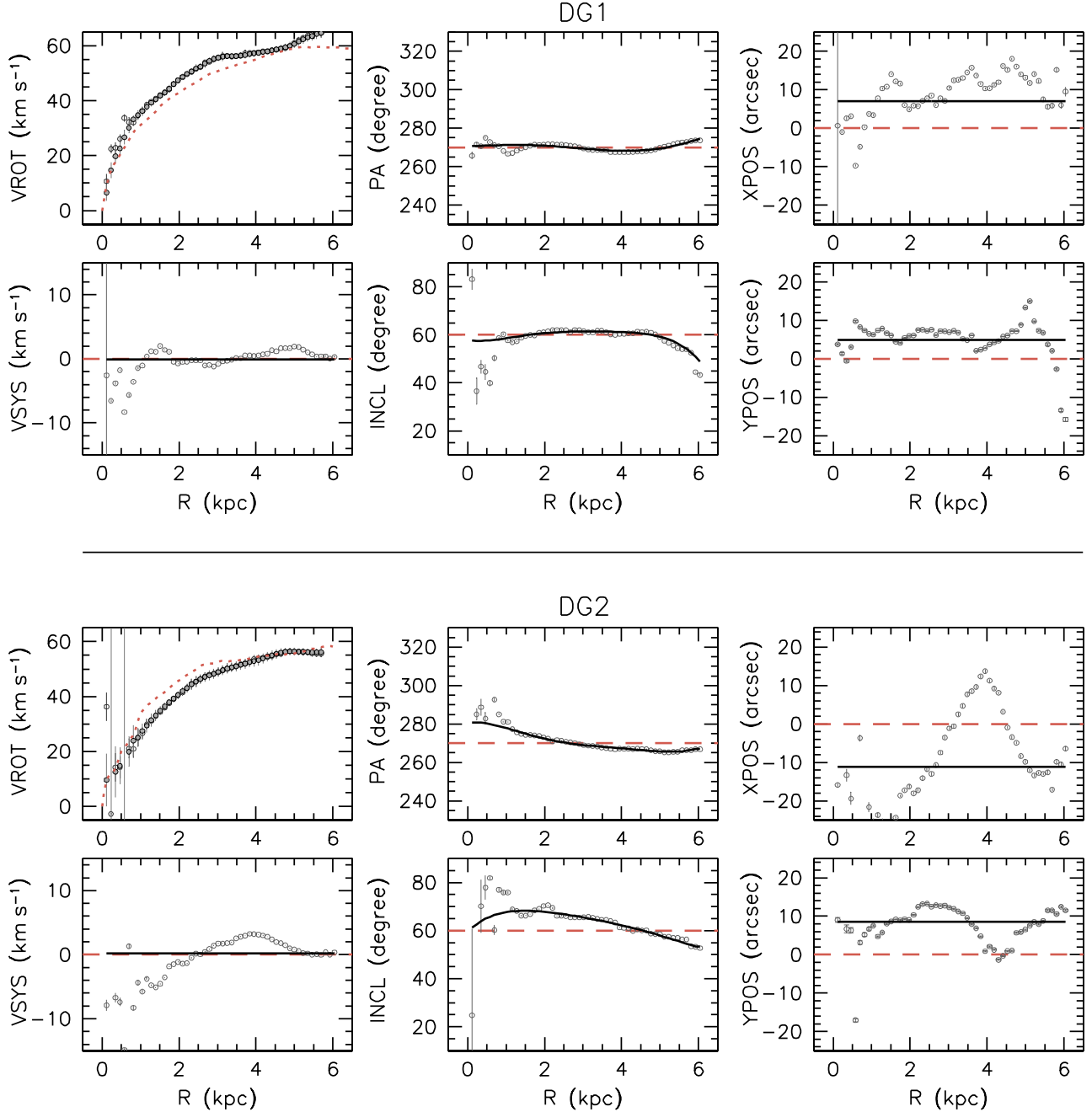}
\caption{\scriptsize The tilted-ring models derived from the hermite $h_{3}$ velocity fields of DG1 (upper) and DG2 (lower).
The open circles in all panels indicate the fit made with all parameters free.
The filled black circles and solid lines in all panels show the finally adopted tilted-ring models
as a function of galaxy radius. The dashed lines indicate the geometrical parameters used when
extracting the rotation velocities (dotted lines in the VROT panels) from the true mass distributions
of the simulated galaxies.
\label{DG1_DG2_rotcur}}
\end{figure*}

\begin{figure*}
\epsscale{1.0}
\includegraphics[angle=0,width=1.0\textwidth,bb=30 175 550 690,clip=]{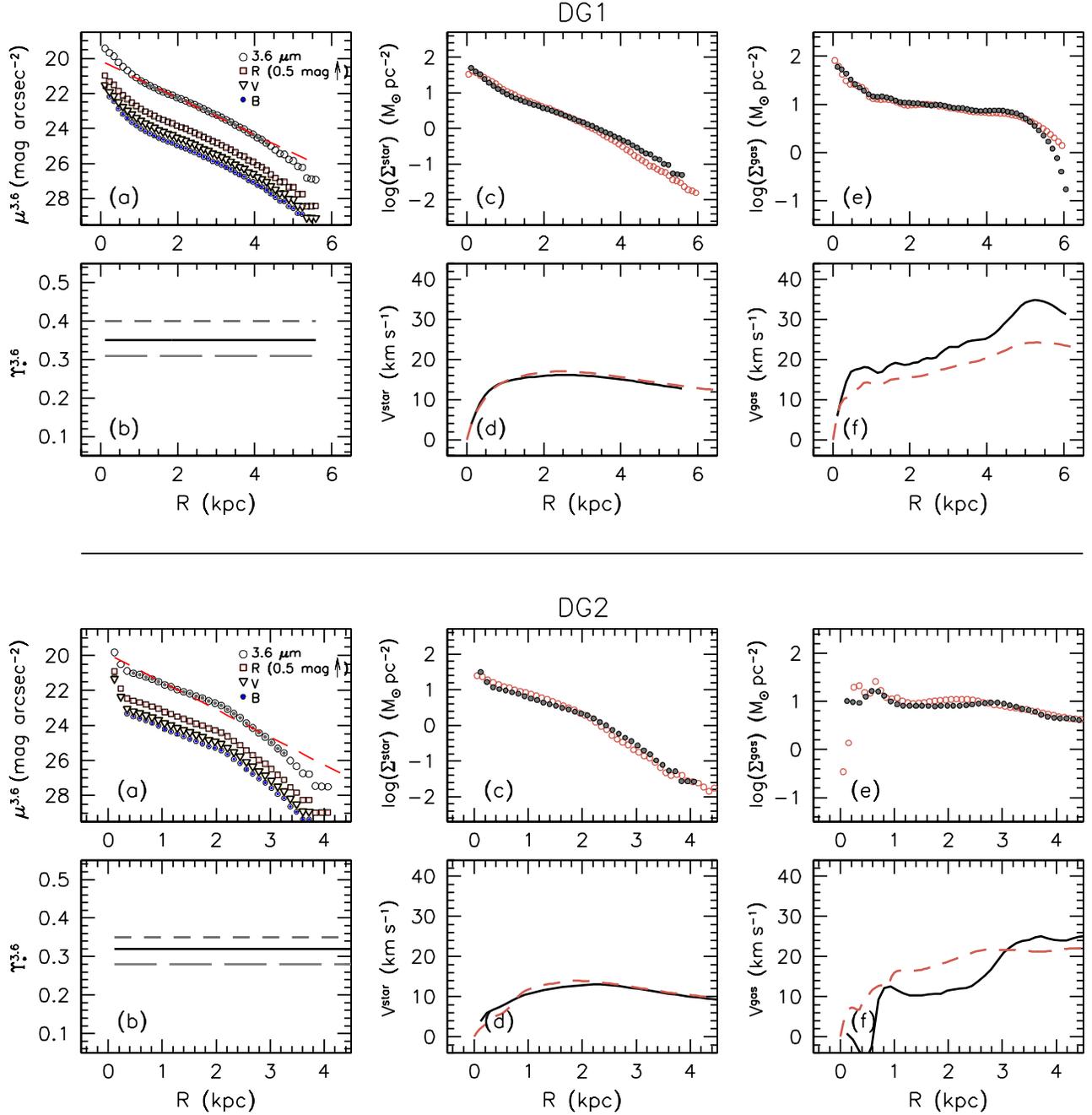}
\caption{\scriptsize Mass models for the baryons of DG1 (top) and DG2 (bottom). 
{\bf (a):} Azimuthally averaged surface brightness profiles (corrected for inclination)
in the simulated 3.6$\mu$m, $R$, $V$, and $B$ bands (top to bottom) derived applying the tilted-ring
parameters shown in Fig.~\ref{DG1_DG2_rotcur}. Note that the $R$-band surface brightness
profile is shifted upward by 0.5 mag for clarity. The dashed line for the 3.6$\mu$m profile
indicates a least-squares fit to the data, the radial range over which the fit is made being indicated by the filled circles.
{\bf (b):} The \ML\ in the 3.6$\mu$m band derived from stellar population
synthesis models. The short and long dashed lines show the \MLsps\ values derived
using optical colors $B-V$ and $B-R$, respectively. The solid line indicates
the mean value adopted as the final \MLsps.
{\bf (c):} The stellar mass surface density derived from the 3.6$\mu$m surface brightness
in (a) using the \MLsps\ value shown in the panel (b).
The red open circles indicate the true profile derived from the simulations. 
{\bf (d):} The rotation velocity for the stellar component derived from the stellar mass density
profile (dots) in the panel (c).
The dashed line shows the true rotation velocity for the stellar component derived from the
true profile (open circles) shown in the panel (c).
{\bf (e):} The radial mass surface density distribution of the gas component scaled by 1.4 to account for He and metals.
The red open circles indicate the true profile derived from the simulations. 
{\bf (f):} The gas rotation velocity derived from the gas surface density profile (dots) in the panel (e).
The dashed line shows the true rotation velocity for the gas component derived from the
true profile (open circles) shown in the panel (e).
\label{DG1_DG2_baryons}}
\end{figure*}

\begin{deluxetable*}{lccccrcrrrr}
\tabletypesize{\scriptsize}
\label{galaxy_properties}
\tablewidth{0pt}
\tablecaption{Properties of the simulated and the THINGS dwarf galaxies}
\tablehead{
\colhead{Name}        & \colhead{$D$}   & \colhead{$\langle \rm Incl. \rangle$} & \colhead{$z_{0}$} & \colhead{$M_{\rm B}$} & \colhead{$\rm V_{\rm max}$} & \colhead{$\rm R_{\rm max}$} & \colhead{$\rm M_{\rm dyn}$} & \colhead{$\rm M_{\rm halo}$} & \colhead{$\rm M_{\rm star}$} & \colhead{$\rm M_{\rm gas}$} \\
                      & \colhead{(Mpc)} & \colhead{($^{\circ}$)}                & \colhead{(kpc)}   & \colhead{(mag)}  & \colhead{(\kms)}  & \colhead{(kpc)} & \colhead{($10^{9}\,M_{\odot}$)}  & \colhead{($10^{9}\,M_{\odot}$)} & \colhead{($10^{8}\,M_{\odot}$)} & \colhead{($10^{8}\,M_{\odot}$)}           \\
                      & \colhead{(1)}   & \colhead{(2)}                         & \colhead{(3)}     & \colhead{(4)}         & \colhead{(5)}                   & \colhead{(6)} & \colhead{(7)} & \colhead{(8)} & \colhead{(9)} & \colhead{(10)}}
\startdata
IC 2574     & 4.0 &  55.7  & 0.57  & -18.1 & 77.6  & 10.4 & 14.6 & 53.2 & 10.38 & 18.63 \\
NGC 2366    & 3.4 &  39.8  & 0.34  & -17.2 & 57.5  & 5.6  & 4.3  & 76.9 & 2.58  & 6.98  \\
Holmberg I  & 3.8 &  13.9  & 0.55  & -14.8 & 38.0  & 1.5  & 0.5  & 33.1 & 1.25  & 2.06  \\
Holmberg II & 3.4 &  49.6  & 0.28  & -16.9 & 35.5  & 7.1  & 2.1  & 4.3  & 2.00  & 7.41  \\
M81 dwB     & 5.3 &  44.8  & 0.09  & -14.2 & 39.8  & 0.8  & 0.3  & 870.9 & 0.30  & 0.31  \\
DDO 53      & 3.6 &  27.0  & 0.14  & -13.4 & 32.4  & 2.0  & 0.5  & 2.1  & 0.18  & 0.85  \\
DDO 154     & 4.3 &  66.0  & 0.20  & -14.2 & 53.2  & 8.2  & 5.4  & 2.4  & 0.26  & 3.58  \\ \\
\hline                                                                                 \\
DG1         & 4.0 &  60.0  & 0.41  & -15.9 & 60.0  & 5.0  & 4.2  & 35.0 & 1.81 & 8.04 \\
DG2         & 4.0 &  60.0  & 0.28  & -15.6 & 56.4  & 5.0  & 3.7  & 20.0 & 0.80 & 5.92 \\ 
\enddata
\tablecomments{
{\bf (1):} Distance as given in \cite{Walter_2008}. For DG1 and DG2, we assume they are at a distance of 4 Mpc;
{\bf (2):} Average value of the inclination derived from the tilted-ring analysis;
{\bf (3):} The vertical scale height of disk;
{\bf (4):} Absolute B magnitude as given in \cite{Walter_2008};
{\bf (5):} Maximum rotation velocity;
{\bf (6):} The radius where the rotation velocity $\rm V_{\rm max}$ of the flat part of the rotation curve is measured;
{\bf (7):} Dynamical mass from the measured $\rm V_{\rm max}$ and $\rm R_{\rm max}$.
{\bf (8):} Halo mass $M_{\rm 200}$ determined from $V_{\rm 200}$ using Eq.~\ref{eq:3}. For the THINGS dwarf galaxies
we use the $V_{\rm 200}$ values fitted using $c$ fixed to 5. See Section~\ref{MM_relation} for more details.
For DG1 and DG2, we show the virial mass measured within the virial radius R$_{\rm vir}$ enclosing an
overdensity of 100 times the cosmological critical density;
{\bf (9):} Stellar mass derived in Section~\ref{MD_baryons}. The stellar mass of DDO 154 is from \cite{deBlok_2008};
{\bf (10):} Gas mass derived in Section~\ref{MD_baryons}. The gas mass of DDO 154 is from \cite{Walter_2008}.}
\end{deluxetable*}

\section{The mass modeling of the simulated dwarf galaxies}\label{Cosmic_HDS}
\subsection{The rotation curves}\label{RC}

We first construct the data cubes of DG1 and DG2 by tracing the motions of the gas
component. The beam and velocity resolutions of the cubes are $\sim$6\arcsec\ (corresponding
to 100 pc at a distance of 4 Mpc) and 2.0 \kms, respectively. 
As an example, we show the integrated gas map, velocity field
and velocity dispersion map extracted from the cube for DG1 in Fig.~\ref{DG1_images}.
For the velocity field, we use the Gauss-Hermite polynomial to model the skewness of
a non-Gaussian profile caused by multiple velocity components (\citeauthor{van_der_Marel1993} \citeyear{van_der_Marel1993}).
This function includes an extra parameter, called $h_{3}$, that measures the skewness of
the Gaussian function, and thus provides more reliable central velocities even for
profiles with significant asymmetries. As discussed in \cite{Oh_2011} (see also \citeauthor{deBlok_2008} \citeyear{deBlok_2008}),
the hermite $h_{3}$ velocity field gives a robust estimate for the underlying circular rotation
of a galaxy in which non-circular motions are insignificant, like DG1 and DG2.
Hermite $h_{3}$ polynomials have also been used to extract the velocity fields of the THINGS galaxies sample
(\citeauthor{deBlok_2008} \citeyear{deBlok_2008}).

\cite{Oh_2011} use the bulk velocity fields when deriving the rotation curves of the
THINGS dwarf galaxies, except for DDO 154 and M81dwB in which non-circular motions are insignificant.
Compared to other types of velocity fields (e.g., intensity-weighted mean, hermite $h_{3}$, single Gaussian and peak velocity fields),
the bulk velocity field more effectively minimizes the effect of small-scale random motions
on the derived kinematics of a galaxy (see \citeauthor{Oh_2008} \citeyear{Oh_2008} for details).
However, for a galaxy that is not significantly affected by non-circular motions like DG1 and DG2,
the bulk velocity field is nearly identical to the hermite $h_{3}$ velocity field.

In Fig.~\ref{DG1_images}b, the iso-velocity contours of the extracted hermite $h_{3}$ velocity field
are distorted in some regions, suggestive of non-circular motions. These are mainly due to
the SN-driven gas outflows in the simulations.
However, despite the presence of non-circular motions the overall pattern of the galaxy rotation
is well recovered in the velocity field.

It is customary to use a set of concentric tilted-rings to model
the velocity field of a galaxy, each with its own kinematic center (XPOS, YPOS),
inclination INCL, position angle PA, expansion velocity VEXP, systemic velocity VSYS
and rotation velocity VROT (\citeauthor{Begeman_1989} \citeyear{Begeman_1989}).
PA is the angle measured counter-clockwise from the north direction in the sky
to the major axis of the receding half of the galaxy.

\begin{figure*}
\includegraphics[angle=0,width=1.0\textwidth,bb=20 305 570 695,clip=]{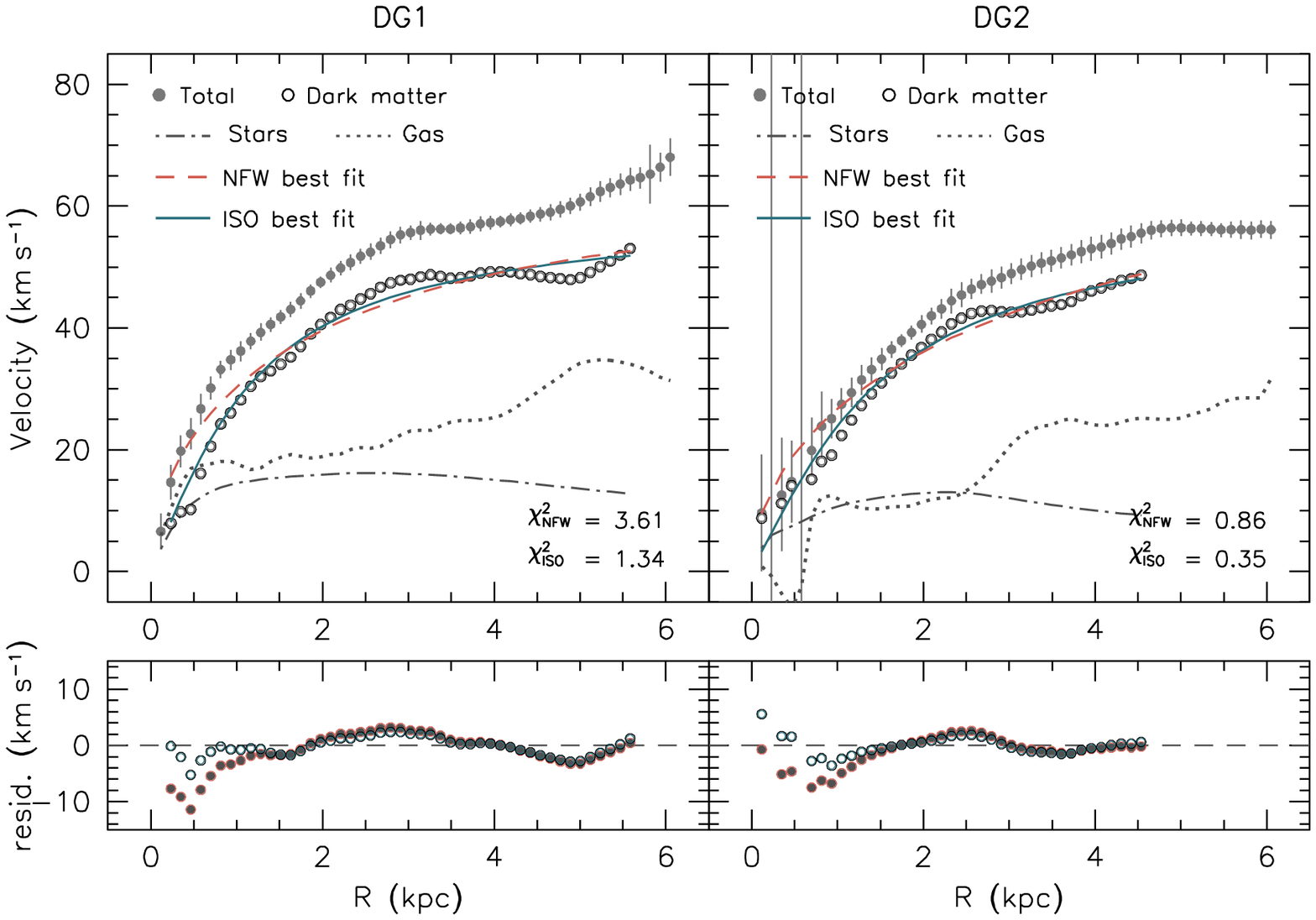}
\caption{\scriptsize The Disk-halo decompositions of the rotation curves of DG1 (left) and DG2 (right).
The gray dots represent the total rotation curves derived from the hermite $h_{3}$ velocity fields.
The dotted and dash-dotted lines indicate the rotation curves of the gas and stellar components, respectively.
The open circles show the dark matter rotation curves derived subtracting the rotation curves of baryons
from the total rotation curves. The dashed and solid lines show the best fitted NFW and pseudo-isothermal
halo models to the dark matter rotation curves, respectively. The reduced $\chi^{2}$ value for each halo model
is denoted on the panels. The lower panels show the velocity residuals between the dark matter rotation curves
and the best fitted halo models. The dots and open circles indicate the results from the NFW and pseudo-isothermal
halo models, respectively.
\label{DH_decomp}}
\end{figure*}

Assuming an infinitely thin disk, we fit these tilted-rings to the hermite $h_{3}$
velocity fields of DG1 and DG2 to derive their rotation curves. 
The derived tilted-ring models for DG1 and DG2 are shown in Fig.~\ref{DG1_DG2_rotcur}.
The errorbar in the rotation velocities indicates the dispersion of individual
velocity values found along a tilted-ring. 
As mentioned earlier, gas outflows driven by SN explosions in the simulations cause non-circular motions
at some regions and induce larger uncertainties in the fitted tilted-rings.
These are seen as the scatter of the very first run results (open circles) with
all ring parameters free in Fig.~\ref{DG1_DG2_rotcur}. However, the local scatter
averages out after several iterations and the final rotation curves (solid lines)
in Fig.~\ref{DG1_DG2_rotcur} seem to give a good description of the underlying
kinematics of DG1 and DG2. As shown in the VROT panels of Fig.~\ref{DG1_DG2_rotcur}, this can be confirmed
by the true rotation velocities (dotted lines) derived using the full three dimensional
mass distributions rather than the projected two dimensional surface density profiles
of the simulated galaxies. Despite not only the uncertainties but also the
assumption of the tilted-ring analysis, i.e., an infinitely thin disk, the difference
between the true and derived rotation velocities is less than 5 \kms, equivalent to about
twice the velocity resolution of the cubes.

For galaxies whose velocity dispersions are large enough compared to their maximum rotation
velocities, we need to correct for the asymmetric drift to obtain more reliable rotation velocities
(\citeauthor{Bureau_2002} \citeyear{Bureau_2002}). However, the $2^{\rm{nd}}$ moment maps of DG1 and DG2
show small velocity dispersions ($\sim$7 \kms) compared to the maximum rotation velocities ($\sim$60 \kms),
and the pressure support is insignificant with respect to the circular rotations.
Therefore, we ignore the asymmetric drift corrections for DG1 and DG2.

\subsection{The mass models of baryons}\label{MD_baryons}

The derived rotation curves in the previous section represent the total kinematics
of the galaxies, including not only dark matter but also the stellar
and gas components. We therefore construct mass models of the baryons
and subtract them from the total kinematics to separate the dark matter component only.

For the gas component, we first derive the gas surface density profile by applying the
derived tilted-rings in Section~\ref{RC} to the integrated gas map shown in the upper-left panel of Fig.~\ref{DG1_images}.
The derived gas surface density profile is then scaled by a factor of 1.4 to account for
helium and metals. The resulting gas surface density profiles of DG1 and DG2 are given in
panels (e) of Fig.~\ref{DG1_DG2_baryons}. From these, we calculate the rotation
velocity due to the gas component, assuming an infinitely thin disk. The derived gas rotation
velocities of DG1 and DG2 are shown in panels (f) of Fig.~\ref{DG1_DG2_baryons}. We also overplot
the true values derived using the full three dimensional mass distribution of the gas components
of DG1 and DG2 as shown in the open circles in the panels (e) of Fig.~\ref{DG1_DG2_baryons}.
For DG1, the true and derived gas surface density profiles are similar but the derived rotation
velocity is systematically higher than the true one. This can be due to the assumption
of ``an infinitely thin disk\footnote[1]{The gas rotation velocities assuming ``an exponential density law''
with scale heights in the range of 0.5--2.5 kpc are somewhat similar to the true one but
slightly higher in the outer regions, possibly due to SNe-driven gas outflows or flaring.}''
which makes one overestimate the gas rotation velocity of a galaxy
with a considerable gas thickness. However, the velocity difference ($\sim$5 \kms) is not very significant.
For DG2, the true gas velocity is systematically higher in the range 0--3\,kpc, particularly in the inner region.
Likewise, the true surface density profile is higher than the derived one in the range 0--3\,kpc.
As shown in Fig.~\ref{DG1_DG2_rotcur}, this is partially because the smaller inclination value
(60$^{\circ}$) is used for extracting the true gas surface density profile from the simulation in
the inner region of DG2. In addition this may also be due to significant vertical gas outflows
or flaring driven by SNe perpendicular to the disk of the galaxy.

Similar to the gas component, we derive the surface density profiles for the stellar components of DG1 and DG2.
For this, as shown in Fig.~\ref{DG1_images}, we use the simulated $B$, $V$, $R$ and {\it Spitzer} IRAC1 3.6$\mu$m
images. In particular, the {\it Spitzer} IRAC1 3.6$\mu$m image is
useful for tracing the underlying old stars which are
usually the dominant stellar population in dwarf galaxies. For the same reason, {\it Spitzer} IRAC 3.6$\mu$m
images have also been used for making mass models for the stellar components of the THINGS galaxies sample
(\citeauthor{deBlok_2008} \citeyear{deBlok_2008}; \citeauthor{Oh_2008} \citeyear{Oh_2008}, \citeyear{Oh_2011}). 

We derive the surface brightness profiles of DG1 and DG2 by applying the tilted-rings derived
in Section~\ref{RC} to their simulated $B$, $V$ and $R$ as well as 3.6$\mu$m images. Both DG1 and DG2 are bulgeless
as found from fitting a Sersic profile to their {\it i} band images (\citeauthor{Governato_2010} \citeyear{Governato_2010}).
To convert the surface brightness profiles to the mass density profiles in units of \surfdens, we obtain the 3.6$\mu$m mass-to-light
(\MLsps) values using an empirical relation between \MLsps\ and optical colors based on the
\cite{Bruzual_Charlot_2003} stellar population synthesis models (\citeauthor{Oh_2008} \citeyear{Oh_2008};
see also \citeauthor{Bell_2001} \citeyear{Bell_2001}).
The \MLsps\ values used for DG1 and DG2 are shown as the solid lines in panels (b) of Fig.~\ref{DG1_DG2_baryons},
and the resulting stellar surface density profiles are given in panels (c) of Fig.~\ref{DG1_DG2_baryons}.
The true surface density profiles for the stellar components of DG1 and DG2 are also overplotted as the
open circles in Fig.~\ref{DG1_DG2_baryons}, and they are similar to the derived ones. This can be
treated as circumstantial evidence that the assumption used for the stellar distribution and \MLsps\ values
provide a good description for the stellar components of DG1 and DG2.

From these, we then compute the corresponding stellar rotation velocities assuming a vertical $\rm {sech}^{2}({\it z})$
scale height distribution of stars. We calculate the vertical scale height $z_{0}$ using a ratio of $h/z_{0}$=$2.5$
where $h$ is the radial scale length of disk, derived from the 3.6$\mu$m surface brightness profile. 
The derived scale heights $z_{0}$ of DG1 and DG2 are 0.41 kpc and 0.28 kpc, respectively. These are similar
to the mean value (0.32 kpc) of the 7 THINGS dwarf galaxies as given in \cite{Oh_2011}. The derived stellar rotation velocities of DG1 and
DG2 are shown in panels (d) of Fig.~\ref{DG1_DG2_baryons}, and they agree well with the true ones indicated by dashed lines. 

\subsection{The disk-halo decomposition}\label{MD_darkmatter}
We separate the dark matter components of DG1 and DG2 by subtracting the mass models of baryons derived in
Section~\ref{MD_baryons} from their total rotation curves. We then fit two halo models, the NFW and
pseudo-isothermal (ISO) halo models, to these kinematic residuals in order to examine the dark matter distribution
in a quantitative way. The NFW and pseudo-isothermal halo models represent cusp--like and constant density (core) matter distributions
at the centers of galaxies, respectively.

The NFW halo model (\citeauthor{NFW_1996} \citeyear{NFW_1996}, \citeyear{NFW_1997}) is given as,
\begin{equation}
\label{eq:1}
V_{\rm{NFW}}(R) = V_{200}\sqrt{\frac{{\rm{ln}}(1+cx)-cx/(1+cx)}{x[{\rm{ln}}(1+c)-c/(1+c)]}},
\end{equation}
\noindent where $c$ is the parameter quantifying the degree of concentration of the dark matter halo.
$V_{200}$ is the rotation velocity at radius $R_{200}$ where the density contrast with the critical density of the Universe exceeds 200
and $x$ is defined as $R/R_{200}$.

Likewise, the rotation velocity based on the pseudo-isothermal halo model is as follows,
\begin{equation}
\label{eq:2}
V_{\rm{ISO}}(R) = \sqrt{4\pi G\rho_{0}R_C^{2}\Biggl[1-\frac{R_C}{R}{\rm{atan}}\Biggl(\frac{R}{R_C}\Biggr)\Biggr]},
\end{equation}
\noindent where $\rho_{0}$ and $R_C$ are the core-density and core-radius of a halo, respectively.

By comparing the fit qualities of these two halo models to the kinematic residuals for
the dark matter component, we examine which halo model is preferred to describe the derived
dark matter distributions of DG1 and DG2.

As shown in the upper panels of Fig.~\ref{DH_decomp}, compared to the CDM NFW halo model,
despite its feasible fits, the pseudo-isothermal halo model gives a better description
for the derived dark matter distributions of both DG1 and DG2 in terms of the fit quality
(i.e., reduced $\chi^{2}$ values). This is also confirmed by the velocity residuals between
the dark matter rotation curves and the best fitted halo models as shown in the lower panels
of Fig.~\ref{DH_decomp}. The best fitted NFW halo models are too steep to match the inner
regions of the dark matter rotation curves of DG1 and DG2.

\section{The simulations vs. THINGS}\label{Sim_THINGS}
In this section, we compare the derived dark matter distribution of DG1 and DG2
with that of the 7 THINGS dwarf galaxies.
THINGS is a high spectral ($\leq$ 5.2 \kms) and angular ($\sim$6\arcsec) resolution 
H{\sc i} survey for 34 nearby galaxies undertaken
using the NRAO\footnote[2]{The National
Radio Astronomy Observatory is a facility of the National Science
Foundation operated under cooperative agreement by Associated
Universities, Inc.} Very Large Array (VLA) (\citeauthor{Walter_2008} \citeyear{Walter_2008}).
THINGS is complemented with other data, such as from the {\it Spitzer} SINGS survey and ancillary
optical $B$, $V$ and $R$ images taken with the KPNO 2.1m telescope (\citeauthor{Kennicutt_2003} \citeyear{Kennicutt_2003}).
These high-quality multi-wavelength data significantly reduce observational uncertainties and thus
enable us to derive more reliable mass models of the galaxies.

\cite{Oh_2011} performed the dark matter mass modeling of 7 dwarf galaxies
in exactly the same way as for DG1 and DG2 as described in Section~\ref{Cosmic_HDS}.
Basic properties of the galaxies are listed in Table 1. In particular, the selected
7 THINGS dwarf galaxies have similar observational properties as DG1 and DG2, such as
resolution ($\sim$150 pc at the distance of $\sim$4 Mpc), maximum rotation velocity ($<$ 80 \kms),
dynamical mass ($\sim$$10^{9}$ $\ensuremath{M_{\odot}}$) and scale height ($\sim$0.3 kpc).
In addition the THINGS dwarf galaxies are not satellites and they have most likely only
weakly interacted with larger systems, as are the cases of DG1 and DG2.
Therefore, this allows us to make a direct comparison between the simulations and observations,
and examine if the simulated dwarf galaxies are realistic compared to the dwarf
galaxies in the local universe.

\subsection{The relation between $M_{\rm star}$ and $M_{\rm halo}$}\label{MM_relation}

By comparing the stellar masses of DG1 and DG2 to their halo masses, we examine whether
the star formation efficiency of DG1 and DG2 is comparable with that of real galaxies
which have similar dynamical masses. This is important as the number of stars which form
in small halos can put strong constraints on  baryonic feedback and its effects.

It has been found that galaxies of smaller or larger halo masses with respect to the
Milky Way appear to have inefficient star formation (\citeauthor{Navarro_Steinmetz_2000} \citeyear{Navarro_Steinmetz_2000};
\citeauthor{Governato_2007} \citeyear{Governato_2007};
\citeauthor{Li_White_2009} \citeyear{Li_White_2009}).
The low star formation efficiency in high mass galaxies is often attributed to AGN feedback
(\citeauthor{Ciotti_Ostriker_2001} \citeyear{Ciotti_Ostriker_2001}; \citeauthor{Benson_2003} \citeyear{Benson_2003}), while the effects
of supernovae are often invoked to explain the low star formation efficiency in low mass
galaxies (\citeauthor{Larson_1974} \citeyear{Larson_1974}; \citeauthor{White_Rees_1978} \citeyear{White_Rees_1978}).
Recently, \cite{Guo_2010} derived the relation between halo mass and stellar mass from abundance matching by combining
the stellar mass function from the Sloan Digital Sky Survey data release 7
(SDSS/DR7; \citeauthor{Li_White_2009} \citeyear{Li_White_2009}) with the halo/subhalo mass function
from N-body \LCDM\ simulations (Millenium and Millenium-II simulations;
\citeauthor{Springel_2005} \citeyear{Springel_2005};
\citeauthor{Boylan-Kolchin_2009} \citeyear{Boylan-Kolchin_2009};
see also \citeauthor{DeLucia_2007} \citeyear{DeLucia_2007} based on semi-analytic models).
Similar methods and results are found in \cite{Conroy_2009}, \cite{Moster_2010} and \cite{Trujillo-Gomez_2010}.

\begin{figure}
\includegraphics[angle=0,width=0.5\textwidth,bb=30 183 540 663,clip=]{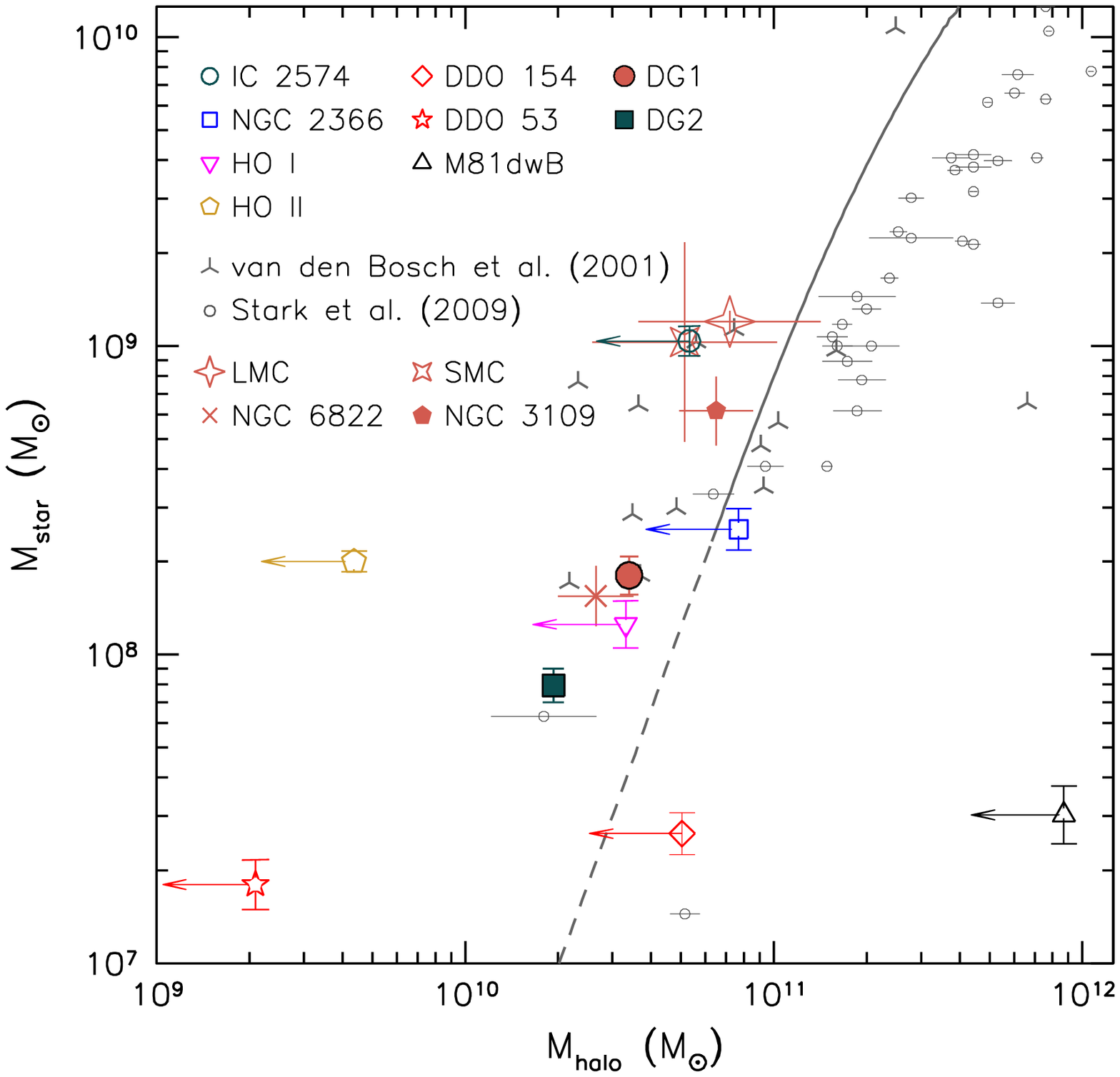}
\caption{\scriptsize The $M_{\rm star}$$\--$$M_{\rm halo}$ relation of DG1, DG2 and the THINGS dwarf galaxies (\citeauthor{Oh_2011} \citeyear{Oh_2011}) as well
as galaxies from the literature (\citeauthor{van_den_Bosch_2001b} \citeyear{van_den_Bosch_2001b};
\citeauthor{Stark_2009} \citeyear{Stark_2009}). The Local Group galaxies,
LMC (\citeauthor{Mastropietro_2005} \citeyear{Mastropietro_2005}; \citeauthor{Guo_2010} \citeyear{Guo_2010}),
SMC (\citeauthor{Stanimirovic_2004} \citeyear{Stanimirovic_2004}; \citeauthor{Guo_2010} \citeyear{Guo_2010}),
NGC 6822 (\citeauthor{Valenzuela_2007} \citeyear{Valenzuela_2007}) and 
NGC 3109 (\citeauthor{Valenzuela_2007} \citeyear{Valenzuela_2007}) are also overplotted.
The errorbars for $\rm M_{halo}$ and $\rm M_{star}$ of the LMC and SMC are computed based on
the different estimates given by \cite{Guo_2010} and the other two papers (i.e., \citeauthor{Mastropietro_2005} \citeyear{Mastropietro_2005}
and \citeauthor{Stanimirovic_2004} \citeyear{Stanimirovic_2004}). The errorbars for NGC 6822 and NGC 3109 come from
the different mass models that \cite{Valenzuela_2007} considered for these galaxies to reproduce their
kinematical and photometrical properties assuming that they are hosted in CDM halos.  
The solid curve is from abundance matching by combining the stellar mass function from the SDSS/DR7 with
the halo/subhalo mass function from the Millenium and Millenium-II simulations (\citeauthor{Guo_2010} \citeyear{Guo_2010}).
As described in \cite{Guo_2010}, the relation below the stellar mass $10^{8.3}$ was extrapolated
assuming constant slope as indicated by the dashed line. See Section~\ref{MM_relation} for more details.
\label{MsMd_relation}}
\end{figure}

\begin{figure*}
\includegraphics[angle=0,width=1.0\textwidth,bb=20 420 585 695,clip=]{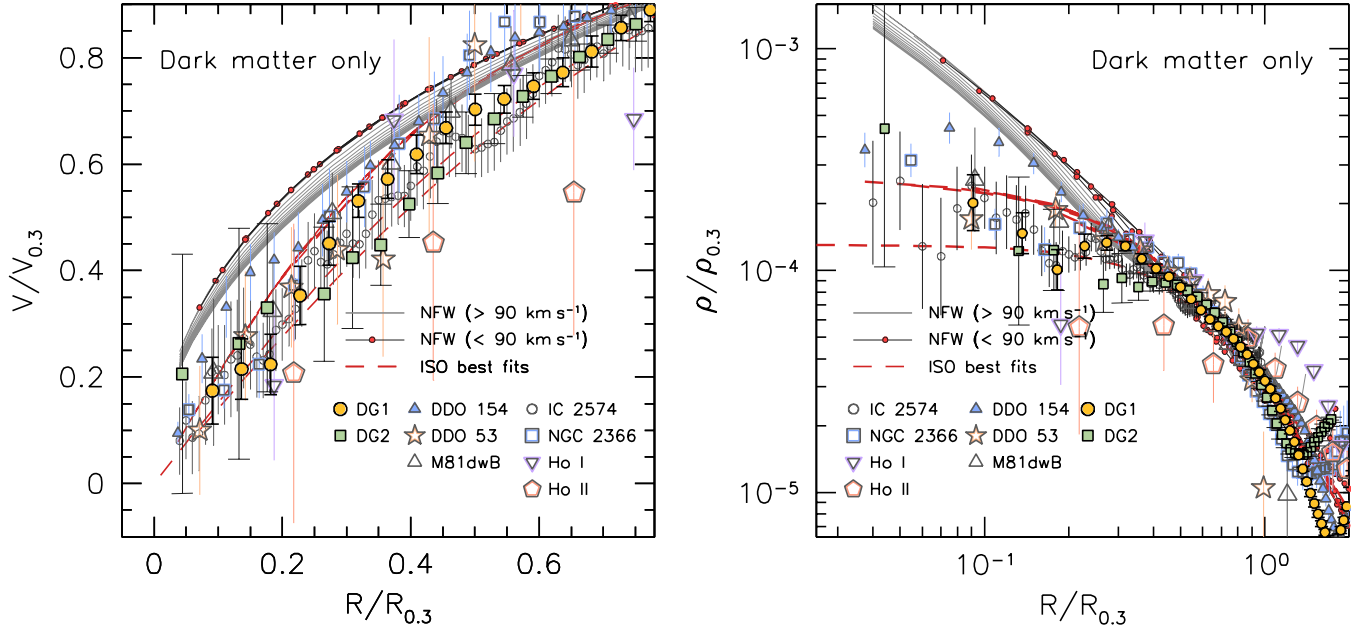}
\caption{\scriptsize {\bf Left:} The rotation curve shape of DG1 and DG2 as well as the 7 THINGS dwarf galaxies.
The dark matter rotation curves (corrected for baryons as shown in Fig.~\ref{DH_decomp})
are scaled with respect to the rotation velocity $V_{0.3}$ at $R_{0.3}$ where the logarithmic slope
of the curve is $d{\rm log}V/d{\rm log}R=0.3$ (\citeauthor{Hayashi_2006} \citeyear{Hayashi_2006}).
The small dots indicate the NFW model rotation curves with $V_{200}$ ranging from 10 to 90 \kms. See text for further details.
The best fitted pseudo-isothermal halo models (denoted as ISO) are also overplotted. See Section~\ref{RC_shape} for more details.
{\bf Right:} The scaled dark matter density profiles of DG1 and DG2 as well as the 7 THINGS dwarf galaxies.
The profiles are derived using the scaled dark matter rotation curves in the left panel.
The small dots represent the NFW models ($\alpha$$\sim$$-1.0$) with $V_{200}$ ranging from 10 to 90 \kms.
The dashed lines indicate the best fitted pseudo-isothermal halo models ($\alpha$$\sim$$0.0$).
See Section~\ref{scaled_DM_profiles} for more details.
\label{scaled_Vrots_DM_profiles}}
\end{figure*}

\cite{Guo_2010} have shown that the star formation efficiency in most recent hydrodynamical,
cosmological galaxy formation simulations is higher than
that predicted from the $M_{\rm star}$$\--$$M_{\rm halo}$ relation.
In particular, as shown in Fig.~\ref{MsMd_relation}, the stellar masses of DG1 and DG2
are about an order of magnitude larger than those inferred from the relation.

However, since the SDSS/DR7 only covers the stellar mass range from $10^{8.3}$ to $10^{11.8}$,
the $M_{\rm star}$$\--$$M_{\rm halo}$ relation outside this range was extrapolated assuming
a constant slope as indicated by the dashed line as shown in Fig.~\ref{MsMd_relation} (see \citeauthor{Guo_2010} \citeyear{Guo_2010}).
The stellar masses of DG1 and DG2 fall within the lower extrapolated region.
Here, we make a more direct comparison between the stellar mass to halo mass ratio between
simulated and observed galaxies. 
To this end, we compare DG1 and DG2 with nearby low-mass galaxies from \cite{van_den_Bosch_2001b}, \cite{Stark_2009} and
the local group as well as the 7 dwarfs from THINGS \citeauthor{Oh_2011} \citeyear{Oh_2011}).

For the halo mass of the THINGS dwarf galaxies, we estimate $M_{\rm 200}$ as follows,
\begin{eqnarray}
\label{eq:3}
M_{\rm{200}}\,[M_{\odot}] &=& 200\times{\frac{3H_{0}^{2}}{8\pi G}}\times{\frac{4\pi R_{200}^3}{3}} \nonumber \\
		&\simeq& 100\times{\frac{H_{0}^{2}}{G}}\times{(\frac{V_{200}}{10H_{0}})^{3}} \nonumber \\
		&\simeq& 3.29\times10^{5}\times V_{200}^{3},
\end{eqnarray}
where $H_{0}$ is the Hubble constant ($70.6\,\kms\,\rm Mpc^{-1}$; \citeauthor{Suyu_2010} \citeyear{Suyu_2010}),
$G$ is the gravitational constant ($4.3\times10^{-3}\,{\rm pc\,M_{\odot}^{-1}}\,{\rm km^{2}\,s^{-2}}$) and
$V_{200}$ in \kms\ is the rotation velocity at radius $R_{200}$ as given in Eq.~\ref{eq:1}.
However, the NFW halo model fails to fit the dark matter rotation curves of the THINGS dwarf
galaxies, giving negative (or close to zero) $c$ values (\citeauthor{Oh_2011} \citeyear{Oh_2011}).
To circumvent the unphysical fits, we instead fit the NFW model to the rotation curves with only $V_{200}$
as a free parameter after fixing $c$ to 5 which is lower than typical values
(e.g., 8--9; \citeauthor{McGaugh_2003} \citeyear{McGaugh_2003}) predicted from \LCDM\
cosmology. The fitted $V_{200}$ values of some galaxies are larger than their measured maximum
rotation velocities. This is because the rotation curves are still rising at the last measured points.
Moreover, as a larger $c$ value induces a smaller $V_{200}$ and hence lower halo mass, our choice of a
low $c$ will provide a robust upper limit for our derived halo mass, as indicated by the arrows in Fig.~\ref{MsMd_relation}.
As shown in Fig.~\ref{MsMd_relation}, despite the uncertainties remaining in these estimates, the stellar
masses of DG1 and DG2 at their given halo masses are consistent with those of real galaxies.
Both the real galaxies and the simulations deviate from the extrapolated line from the $M_{\rm star}$$\--$$M_{\rm halo}$ relation in
\cite{Guo_2010} at low halo masses. However, as discussed in \cite{Trujillo-Gomez_2010}, there still
remain uncertainties for dwarfs in the sense that the observational data suffer from small number statistics and
the results of abundance matching are incomplete in the low-luminosity tail of the luminosity function.

\begin{figure*}
\includegraphics[angle=0,width=1.0\textwidth,bb=20 400 578 695,clip=]{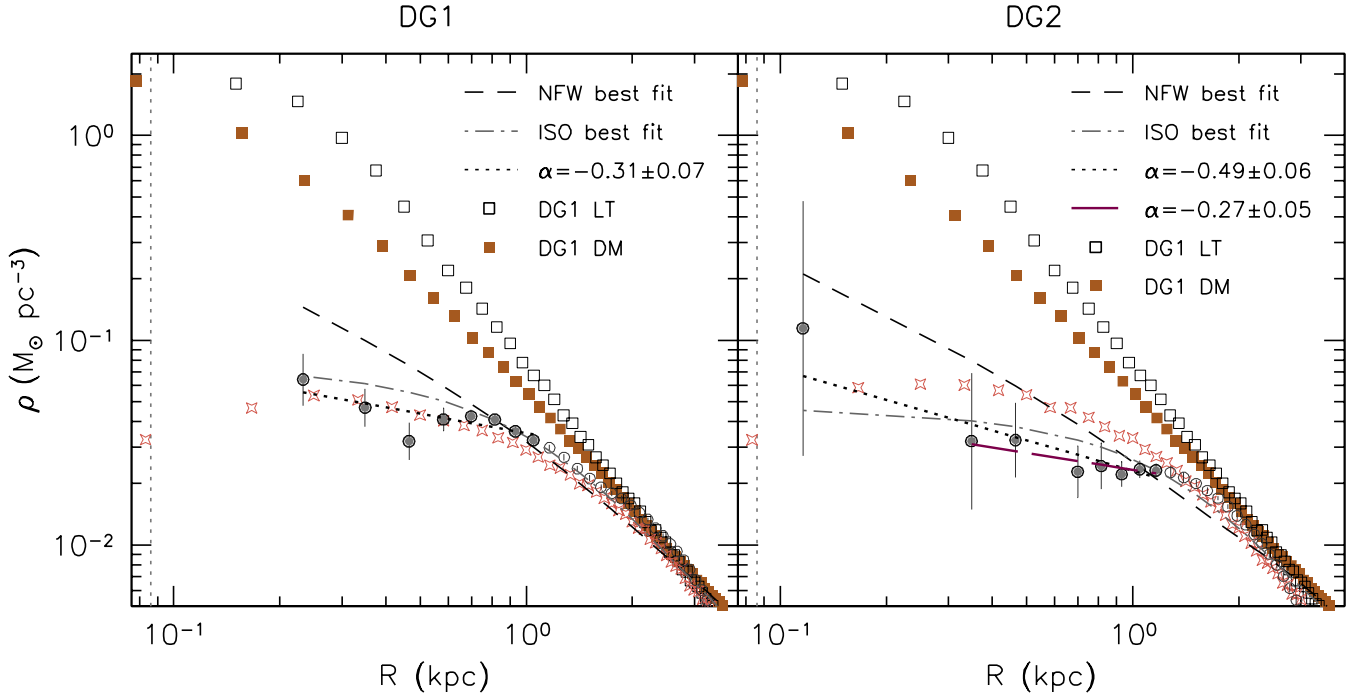}
\caption{\scriptsize The dark matter density profiles of DG1 (left) and DG2 (right).
The circles represent the dark matter density profiles derived from the dark matter rotation curves
shown in Fig.~\ref{DH_decomp}. The short dashed and dash-dotted lines indicate the mass density profiles derived
using the best fitted NFW and pseudo-isothermal halo models in Fig.~\ref{DH_decomp}, respectively.
The open and filled rectangles indicate the density profiles of DG1LT and DG1DM, respectively. See Section~\ref{SPH_description}
for more details.
The inner slope of the profile is measured by a least squares fit (dotted lines) to the data points less than 1.0 kpc as indicated by gray dots.
In the right panel the long dashed line indicates a least squares fit, excluding the innermost point.
The measured inner slope $\alpha$ is shown in the panel. The true dark matter density profiles
in the simulations are also overplotted as indicated by open stars. The vertical gray dotted lines indicate
the force resolution (86 pc) of the simulations.
\label{MD_profiles}}
\end{figure*}

\subsection{The rotation curve shape}\label{RC_shape}
The rotation curve reflects the total potential (dark matter $+$ baryons) of the galaxy
and thus it is directly related to the radial matter distribution in the galaxy (and vice versa).
Consequently, the cusp--like dark matter distributions in the CDM halos
impose a unique shape on the rotation curves, which steeply rise at the inner regions.
Therefore, a relative comparison of galaxy rotation curves between the simulations and observations
can serve as a useful constraint for testing the \LCDM\ simulations.

In this context, we compare the rotation curves of DG1 and DG2 with those of the THINGS dwarf galaxies.
In order to accentuate their inner shapes, we scale the rotation curves of both the simulations
and the THINGS dwarf galaxies with respect to the velocity $V_{0.3}$ at the radius $R_{0.3}$ where the
logarithmic slope of the curve is $d{\rm log}V/d{\rm log}R=0.3$ (\citeauthor{Hayashi_2006} \citeyear{Hayashi_2006}).
At the scaling radius $R_{0.3}$, the rotation curves of both simulations and the observations are well resolved,
which allows any differences between them to show up.

The scaled rotation curves, with the kinematic contribution of baryons subtracted, are shown in the left panel of Fig.~\ref{scaled_Vrots_DM_profiles}.
We overplot the scaled rotation curves of NFW CDM halos (dark-matter-only) with different maximum rotation velocities
ranging from 10 to 350 \kms. We choose c values of $\sim$$9$ and $\sim$$8$ for dwarf and disk galaxies
respectively, which in turn provide $V_{200}$ values ranging from $\sim$$10$ to $\sim$$90$ \kms,
and $\sim$$100$ to $\sim$$350$ \kms, respectively. 
Considering that the rotation velocities of DG1, DG2 and the THINGS dwarf galaxies at the
outermost measured radii are all less than $\sim$80 \kms, the CDM rotation curve
with $V_{200}$$\sim$90 \kms\ (i.e., small dots in Fig.~\ref{scaled_Vrots_DM_profiles}) can be
regarded as a lower limit to the maximum rotation velocities of the galaxies.

As shown in the left panel of Fig.~\ref{scaled_Vrots_DM_profiles}, the scaled rotation curves of the THINGS dwarf galaxies are similar 
to each other, showing a linear increase in the inner regions. The inner shapes of the rotation curves
are better described by pseudo-isothermal halo models (dashed lines) than the NFW models.
This implies that the THINGS dwarf galaxies have core-like dark matter distributions at their
centers (see \citeauthor{Oh_2011} \citeyear{Oh_2011} for more discussion).
Similarly, the scaled rotation curves of DG1 and DG2 are consistent with
those of the THINGS dwarf galaxies. They significantly deviate from the CDM rotation curves
at the inner regions and, like the THINGS dwarf galaxies, they increase
too slowly to match the steep rotation curves of the CDM halos.

\subsection{The dark matter density profile}\label{scaled_DM_profiles}
A more direct way to examine the central matter distributions in galaxies is to convert the galaxy rotation curve
to the mass density profile. In particular, the measurement of the inner slope of the profile provides
a stringent observational constraint on the ``cusp/core'' problem. With an assumption of a spherical
mass distribution for the galaxy halo, the galaxy rotation curve $V(R)$ can be converted to the mass density
profile $\rho(R)$ by the following formula (see \citeauthor{deBlok_2001} \citeyear{deBlok_2001} and
\citeauthor{Oh_2008} \citeyear{deBlok_2008}, \citeyear{Oh_2011} for more details),
\begin{equation}
\label{eq:4}
\rho(R) = \frac{1}{4\pi G}\Biggl[2\frac{V}{R}\frac{\partial V}{\partial R} + \Biggl(\frac{V}{R}\Biggr)^{2}\Biggr],
\label{rho_poisson}
\end{equation}
where $V$ is the rotation velocity observed at radius $R$, and $G$ is the gravitational constant.
Here we do not de-contract the halos since in these galaxies adiabatic contraction does not occur and
rather expansion happens as shown in \cite{Governato_2010} (see also \citeauthor{Dutton_2007} \citeyear{Dutton_2007}).

Using Eq.~\ref{rho_poisson}, we derive the dark matter density profiles of the THINGS dwarf galaxies, DG1 and DG2
as well as the CDM halos whose rotation curves are shown in the left panel of Fig.~\ref{scaled_Vrots_DM_profiles}.
In addition, we also derive the corresponding mass density profiles of the best fitted pseudo-isothermal halo models
to the THINGS dwarf galaxies.
As shown in the right panel of Fig.~\ref{scaled_Vrots_DM_profiles}, despite the scatter, both DG1 and DG2 have shallower 
mass density profiles than dark matter only simulations. Instead, they are more consistent with the THINGS dwarf galaxies
showing near--constant density dark matter distributions at the centers.

In Fig.~\ref{MD_profiles}, we compare the derived dark matter density profiles of DG1 and DG2 with
their true full 3$-$D dark matter density distribution. The inner decrease in the actual dark matter
density profiles of Fig.~\ref{MD_profiles} is due to the shape of the potential in the region below
the force resolution (86 pc). As shown in Fig.~\ref{MD_profiles}, for DG1, the observationally derived dark matter
density profile robustly traces the true values but that for DG2 it is found to be on average a factor of three
lower than its true value at the central regions. This is mainly due to the lower gas rotation velocity of
DG2 as shown in panel (f) of Fig.~\ref{DG1_DG2_baryons}, resulting in smaller velocity gradients ${\partial V}/{\partial R}$
in Eq.~\ref{rho_poisson} and thus smaller densities. However, considering the uncertainties in deriving the
profile, the recovered profile is acceptable to examine the central dark matter distribution.

We determine the inner density slopes $\alpha$ assuming a power law ($\rho \sim r^{\alpha}$) and find them to be $\alpha$$=$$-0.31\pm 0.07$
for DG1 and $\alpha$$=$$-0.49\pm 0.06$ for DG2, respectively. If we re-measure the slope of DG2, excluding the
innermost point which has a large errorbar, the slope is flatter ($\alpha$$=$$-0.27\pm 0.05$) as indicated
by the long dashed line in the right panel of Fig.~\ref{MD_profiles}. 
These slopes deviate from the steep slope of $\sim$$-1.0$ from dark-matter-only cosmological simulations.
The profiles of both DG1 and DG2 deviate from NFW models beyond about 10 times
the force resolution. This tells us that the baryonic feedback processes
in dwarf galaxies can affect the dark matter distribution in such a way that the central cusps predicted from
dark-matter-only simulations are flattened, resulting in dark matter halos characterised by a core, as found in normal dwarf
galaxies in the local universe.

\section{Conclusions}\label{conclusion}
In this paper, we have compared the dark matter distribution of the dwarf galaxies from a novel
set of SPH+N-body simulations by \cite{Governato_2010} with that of 7 THINGS dwarf galaxies
to address the ``cusp/core'' problem in \LCDM.
The simulations were performed in a fully cosmological context, and include
the effect of baryonic feedback processes, particularly strong gas outflows driven by SNe.
Both the simulated and the observed dwarf galaxies have similar kinematic properties,
and have been analyzed in a homogeneous and consistent manner as described in \cite{Oh_2011}.
The techniques used in deriving dark matter density profiles were found to provide accurate
results when compared with the true underlying profiles, supporting the veracity of the
techniques employed by observers. Therefore, this provides a quantitative comparison between
the simulations and the observations, and allows us to examine how the baryonic feedback processes
affect the dark matter distribution at the centers of dwarf galaxies.

From this, we test the general predictions from \LCDM\ simulations:
(1) the steep rotation curve inherent in the central cusp, and (2) the steep inner slope of $\sim$$-1.0$
of the dark matter density profiles. We find that the dark matter rotation curves of the newly
simulated dwarf galaxies rise less steeply at the centers than those from dark-matter-only simulations.
Instead, they are more consistent with those of the THINGS dwarf galaxies. In addition, the mean value
of the inner density slopes $\alpha$ of the simulated dwarf galaxies is $\simeq$$-0.4\pm0.1$. Compared to
the steep slope of $\sim$$-1.0$ predicted from the previous dark-matter-only simulations (including
our simulations run with DM only), these flat slopes are in better agreement with $\alpha$$=$$-0.29\pm 0.07$
found in the 7 THINGS dwarf galaxies analysed by \citet{Oh_2011}.

In conclusion, the results described in this paper confirm that energy transfer and
subsequent gas removal in a clumpy ISM have the net effect of causing
the central DM distribution to expand, while at the same time limiting
the amount of baryons at the galaxy center.  By the present time the DM
central profile in galaxies DG1 and DG2 is well approximated by a
power law with slope $\alpha$ of $\sim$$-0.4\pm0.1$. These values
of $\alpha$ are significantly flatter than in the collisionless
control run and are in agreement with those of observed shallow DM
profiles in nearby dwarf galaxies.

\acknowledgements
SHOH acknowledges financial support from the South African Square
Kilometre Array Project. FG acknowledges support from HST GO-1125, NSF ITR grant PHY-0205413
(also supporting TQ), NSF grant AST-0607819 and NASA ATP
NNX08AG84G. The work of WJGdB is based upon research supported by the South
African Research Chairs Initiative of the Department of Science and
Technology and National Research Foundation. We thank the computer resources and
technical support by TERAGRID, ARSC, NAS and the UW computing center,
where the simulations were run.

\bibliography{THINGS_DGs_AJ_2col_version}

\end{document}